\documentclass[10pt,a4paper,twoside]{article}
\usepackage{epsfig}
\usepackage{baltlat6}
\usepackage{array}
\usepackage{dcolumn}

\begin{document}
\ \
\vspace{-0.5mm}

\setcounter{page}{53}
\vspace{-2mm}

\titlehead{Baltic Astronomy, vol.\,20, 53--63, 2011}

\vspace{-2mm}

\titleb{CHEMICAL COMPOSITION OF THE RS CVn-TYPE STAR \\
33~PISCIUM}

\vspace{-2mm}

\begin{authorl}
\authorb{G.~Barisevi\v{c}ius}{1},
\authorb{G.~Tautvai\v{s}ien\.{e}}{1},
\authorb{S.~Berdyugina}{2},
\authorb{Y.~Chorniy}{1} and
\authorb{I.~Ilyin}{3}
\end{authorl}

\begin{addressl}
\addressb{1}{Institute of Theoretical Physics and Astronomy,  Vilnius
University,\\ Go\v{s}tauto 12, Vilnius, LT-01108, Lithuania}

\addressb{2}{Kiepenheuer Institut f\"ur Sonnenphysik, Sch\"oneckstr. 6,
D-79104 Freiburg,\\ Germany}

\addressb{3}{Astrophysical Institute Potsdam, An der Sternwarte 16,
Potsdam D-14482,\\ Germany}
\end{addressl}

\submitb{Received: 2011 March 7; accepted 2011 March 25}

\begin{summary} Abundances of 22 chemical elements, including the key
elements and isotopes such as $^{12}{\rm C}$, $^{13}{\rm C}$, N and O,
are investigated in the spectrum of 33~Psc, a single-lined RS~CVn-type
binary of low magnetic activity.  The high resolution spectra were
observed on the Nordic Optical Telescope and analyzed with the MARCS
model atmospheres.  The following main parameters have been determined:
$T_{\rm eff}$ = 4750~K, $\log~g$ = 2.8, [Fe/H] = --0.09, [C/Fe] =
--0.04, [N/Fe] = 0.23, [O/Fe] = 0.05, C/N = 2.14, $^{12}$C/$^{13}$C =
30, which show the first-dredge-up mixing signatures and no
extra-mixing.  \end{summary}


\resthead{Chemical composition of the RS CVn-type star 33~Psc}{ G.~
Barisevi\v{c}ius, G.~Tautvai\v{s}ien\.{e}, S.~Berdyugina et al.}

\sectionb{1}{INTRODUCTION}

This is the third paper in a series dedicated to a detailed study of
photospheric abundances in RS CVn stars (Tautvai\v{s}ien\.{e} et al.\
2010; Barisevi\v{c}ius et al.\ 2010, hereafter Papers I and II) with the
main aim to get the carbon isotope $^{12}$C/$^{13}$C and C/N
ratios in these chromospherically active stars.  We plan to investigate
correlations between the abundance alterations of chemical elements in
the atmospheres of these stars and their physical macro parameters, such
as the speed of rotation and the magnetic field.

Here we present results of the analysis of a bright K0\,IIIb single-line
binary 33~Psc (HD~28) of $V = 4.78$~mag (van Leeuwen 2007).  As a binary
with an orbital period of $P=72.93$~days it was recognized by Harper
(1926).  The presence of the emission in its H and K lines of Ca\,{\sc
ii} was detected spectroscopically by Young \& Konigest (1977).  From
the analysis of intensity of the Ca\,{\sc ii} emission in the K line,
Glebocki \& Stawikovski (1979) concluded that the primary is of
1.6~$M_{\odot}$ and the secondary is of 0.7~$M_{\odot}$.  This
evaluation was confirmed by Pourbaix \& Boffin (2003) using the {\it
Hipparcos} intermediate astrometric data:  the mass of the primary was
found to be $1.7\pm 0.4 M_{\odot}$ and of the secondary $0.76\pm
0.11~M_{\odot}$.  This system has not been resolved by speckle
interferometry (Hartkopf et al.\ 2001).

From narrow-band photometry, Hansen \& Kjaergaard (1971) and
Gottlieb \& Bell (1972) have determined that 33~Psc is slightly
metal-deficient ([Fe/H]\,$\approx$\,--0.2).  The high-resolution
spectroscopic metallicity determinations of 33~Psc are slightly
controversial:  McWilliam (1990) has obtained [Fe/H]\,$\approx$\,--0.31,
while the [Fe/H] values obtained by Randich et al.\ (1994) and Morel et
al.\ (2004) are solar.

In papers I and II we investigated $\lambda$~And and 29~Dra which are
among most active RS~CVn stars. 33~Psc by now has the lowest known
magnetic activity among RS~CVn stars (Strassmeier et al.\ 1988).
Results of Einstein X-ray observations of 33~Psc were presented by
Walter (1985).  Its surface flux is $<5 \cdot 10^3$
erg\,cm$^{-2}$\,s$^{-1}$, which is about a factor of three lover than
the flux from the quiet solar corona, and is two orders of magnitude
below the surface fluxes observed in the least-active long-period RS CVn
systems.  Drake et al.\ (1989) for 33~Psc have found quite low upper
limit of the 6 cm radio flux, implying low activity as well.  The
ultraviolet chromosferic emission lines are also weak (Basri et al.\
1985).

For $\lambda$~And and 29~Dra the ratios of $^{12}$C/$^{13}$C were found
to be lower than it is predicted by the first dredge-up theory of the
first-ascent giants.  Therefore it was interesting to check what are the
effects of mixing processes in 33~Psc, which lies on the giant branch
almost at the same luminosity but is much less active than $\lambda$~And
and 29~Dra.

\sectionb{2}{OBSERVATIONS AND THE METHOD OF ANALYSIS}

The spectrum of 33~Psc was observed twice -- in August of 1999 and
September of 2006 on the 2.56~m Nordic Optical Telescope using the SOFIN
echelle spectrograph with the optical camera, which provided a spectral
resolving power of $R \approx 80\,000$, for 26 slightly shifted spectral
orders, each of $\sim 4$~nm, in the spectral region from 500 to 830 nm.
Details of spectral reductions and the method of analysis are
presented in Paper~I.

\vspace{2mm}
%------------------- Fig. 1
\vbox{
\centerline{\psfig{figure=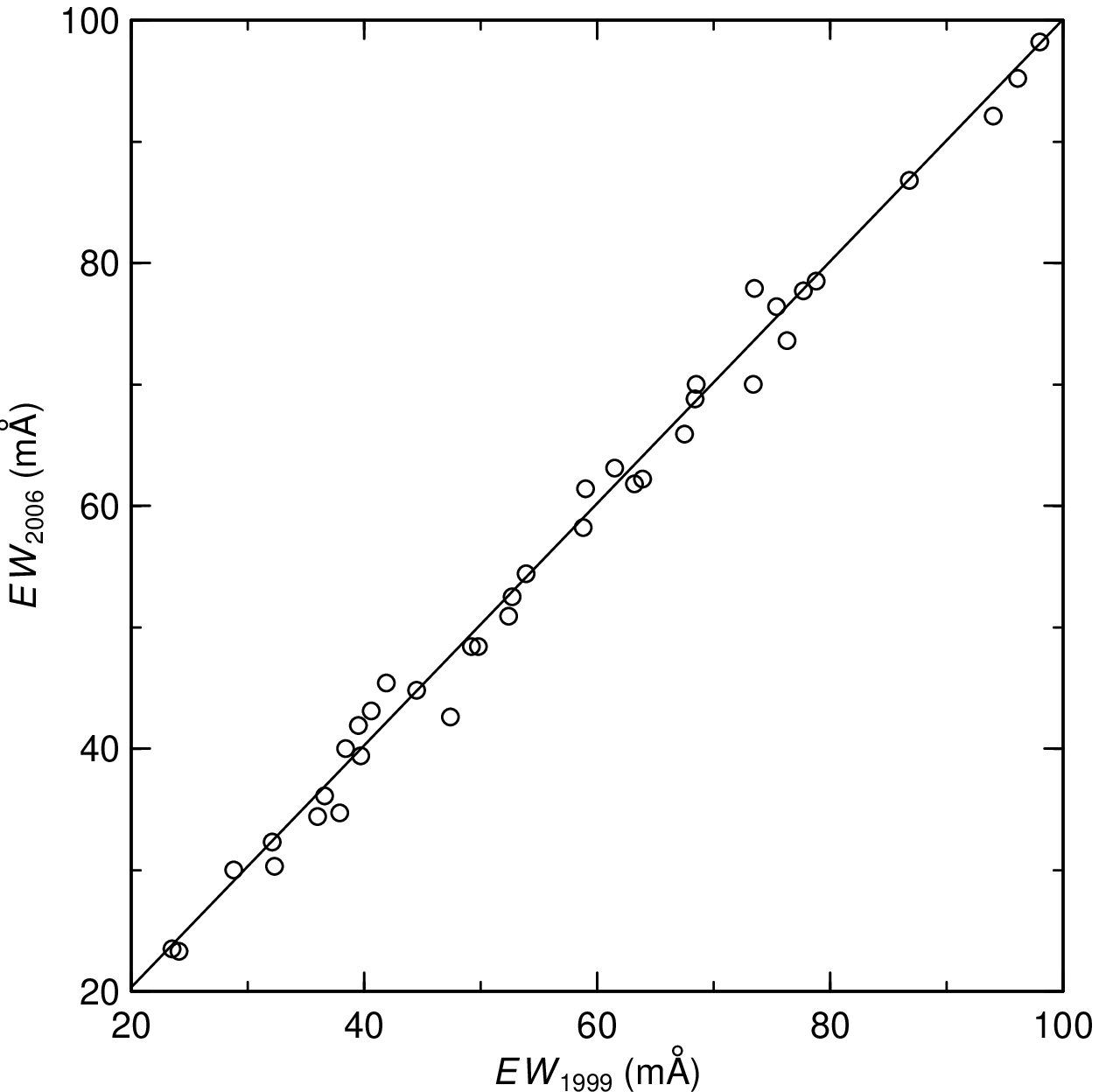,width=80truemm,angle=0,clip=}}
\captionb{1}{A comparison of the equivalent widths measured in
the 33 Psc spectra observed in 1999 and 2006.}
}
\vspace{2mm}

%------------------- Table 1

\begin{table}[!t]
{
\footnotesize
\noindent
\extrarowheight=-.5pt
\tabcolsep=2pt
\begin{tabular}{lcD..{1.1}|lcD..{1.1}|lcD..{1.1}}
\multicolumn{9}{c}{\parbox[c]{120mm}{\baselineskip=4pt
{\smallbf\ \ Table 1.}{\small\ \ Equivalent widths of lines, {\it
EW}, in the spectrum of 33 Psc.
 \lstrut}}}\\
\hline
\multicolumn{1}{l}{Element\hstrut} &
\multicolumn{1}{c}{$\lambda$\,(\AA)} &
\multicolumn{1}{c|}{{\it EW}\,(m\AA)} &
\multicolumn{1}{l}{Element} &
\multicolumn{1}{c}{$\lambda$\,(\AA)} &
\multicolumn{1}{c|}{{\it EW}\,(m\AA)}&
\multicolumn{1}{l}{Element} &
\multicolumn{1}{c}{$\lambda$\,(\AA)} &
\multicolumn{1}{c}{{\it EW}\,(m\AA)\lstrut}\\
\hline
~~Si\,{\sc i} & \hstrut 5517.55~~ &   23.4 & & 6274.66~~ &   74.0 &  & 6392.54$^\star$ &   62.3 \\
	 &    5645.60~~ &   52.0 & 	 &    6285.16~~ &   83.1 & 	 &    6574.21$^\star$ &   93.1 \\
	 &    5793.08~~ &   56.0 & 	 &    6292.82~~ &   87.1 & 	 &    6581.21$^\star$ &   75.9 \\
	 &    5948.54~~ &   91.6 & ~~Cr\,{\sc i}	 &    5712.78~~ &   49.9 & 	 &    6646.97~~ &   43.0 \\
	 &    6131.85~~ &   29.5 & 	 &    5783.87~~ &   76.6 & 	 &    6786.86$^\star$ &   48.8 \\
	 &    7003.57$^\star$ &   51.7 & 	 &    5784.97~~ &   63.2 & 	 &    6793.27$^\star$ &   31.3 \\
~~Ca\,{\sc i}	 &    5260.38~~ &   66.9 & 	 &    5787.92~~ &   81.5 & 	 &    6839.83~~ &   78.7 \\
	 &    5867.57~~ &   53.1 & 	 &    6661.08~~ &   27.5 & 	 &    6842.69~~ &   60.4 \\
	 &    6455.60~~ &   94.5 & 	 &    6979.80~~ &   73.6 & 	 &    6843.65~~ &   79.7 \\
	 &    6798.47$^\star$ &   23.7 & 	 &    6980.91~~ &   27.0 & 	 &    6851.64~~ &   33.3 \\
~~Sc\,{\sc ii}	 &    5526.81~~ &  106.2 & ~~Fe\,{\sc i}	 &    5395.22~~ &   42.3 & 	 &    6857.25~~ &   45.9 \\
	 &    5667.14~~ &   67.5 & 	 &    5406.78~~ &   58.9 & 	 &    6858.15~~ &   71.6 \\
	 &    6279.75~~ &   61.2 & 	 &    5522.45~~ &   65.8 & 	 &    6862.49~~ &   51.0 \\
	 &    6300.69~~ &   22.6 & 	 &    5577.03~~ &   21.1 & 	 &    7461.53$^\star$ &   71.7 \\
~~Ti\,{\sc i}	 &    5648.58~~ &   44.7 & 	 &    5579.35~~ &   23.8 & ~~Fe\,{\sc ii}	 &    5132.68~~ &   33.5 \\
	 &    5662.16~~ &   66.6 & 	 &    5607.67$^\star$ &   36.3 & 	 &    5264.81~~ &   47.0 \\
	 &    5716.45~~ &   33.7 & 	 &    5608.98$^\star$ &   29.4 & 	 &    6113.33~~ &   16.0 \\
	 &    5739.48$^\star$ &   39.6 & 	 &    5651.48~~ &   37.3 & 	 &    6369.46~~ &   23.6 \\
	 &    5880.27~~ &   58.0 & 	 &    5652.33~~ &   48.4 & 	 &    6456.39~~ &   55.2 \\
	 &    5899.30$^\star$ &   95.6 & 	 &    5653.86~~ &   57.6 & 	 &    7711.72$^\star$ &   43.7 \\
	 &    5903.31$^\star$ &   44.7 & 	 &    5679.03~~ &   73.8 & ~~Co\,{\sc i}	 &    5647.23~~ &   60.0 \\
	 &    5941.76~~ &   81.3 & 	 &    5720.90~~ &   33.1 & 	 &    6117.00~~ &   40.8 \\
	 &    5953.17~~ &   87.7 & 	 &    5732.30~~ &   23.9 & 	 &    6188.98~~ &   58.0 \\
	 &    5965.83~~ &   83.1 & 	 &    5738.24$^\star$ &   32.2 & 	 &    6455.00~~ &   44.4 \\
	 &    6064.63$^\star$ &   58.5 & 	 &    5741.86$^\star$ &   52.6 & 	 &    6678.82~~ &   42.4 \\
	 &    6098.66~~ &   25.3 & 	 &    5784.67~~ &   57.8 & ~~Ni\,{\sc i}	 &    5587.87$^\star$ &   98.1 \\
	 &    6121.00~~ &   27.6 & 	 &    5793.92~~ &   56.9 & 	 &    5589.37$^\star$ &   45.0 \\
	 &    6126.22~~ &   83.8 & 	 &    5806.73~~ &   72.6 & 	 &    5593.75$^\star$ &   63.1 \\
	 &    6220.49$^\star$ &   40.7 & 	 &    5807.79~~ &   30.7 & 	 &    5643.09~~ &   30.3 \\
	 &    6303.77~~ &   52.4 & 	 &    5811.92~~ &   27.1 & 	 &    5748.35$^\star$ &   77.7 \\
	 &    6599.11$^\star$ &   66.7 & 	 &    5814.82~~ &   45.7 & 	 &    5805.22~~ &   58.8 \\
	 &    6861.45~~ &   41.9 & 	 &    6027.06~~ &   86.7 & 	 &    6053.68$^\star$ &   36.3 \\
~~V\,{\sc i}	 &    5604.96$^\star$ &   39.2 & 	 &    6034.04~~ &   21.1 & 	 &    6111.08~~ &   54.9 \\
	 &    5646.11~~ &   42.7 & 	 &    6035.35~~ &   17.4 & 	 &    6128.98~~ &   73.8 \\
	 &    5657.45~~ &   52.8 & 	 &    6054.07$^\star$ &   23.5 & 	 &    6130.14~~ &   37.3 \\
	 &    5668.37~~ &   52.8 & 	 &    6056.01$^\star$ &   86.8 & 	 &    6204.60~~ &   43.2 \\
	 &    5743.43$^\star$ &   60.2 & 	 &    6098.25~~ &   32.1 & 	 &    6378.25~~ &   51.4 \\
	 &    6039.74$^\star$ &   69.2 & 	 &    6105.13~~ &   24.0 & 	 &    6598.60$^\star$ &   41.8 \\
	 &    6058.18$^\star$ &   35.2 & 	 &    6120.24~~ &   47.4 & 	 &    6635.13~~ &   44.9 \\
	 &    6111.65~~ &   75.2 & 	 &    6187.99~~ &   75.6 & 	 &    6767.78~~ &  132.3 \\
	 &    6119.53~~ &   82.9 & 	 &    6200.32~~ &  114.1 & 	 &    6772.32$^\star$ &   74.9 \\
	 &    6135.37~~ &   69.8 & 	 &    6226.74$^\star$ &   54.2 & 	 &    6842.03~~ &   51.6 \\
	 &    6224.50$^\star$ &   68.6 & 	 &    6229.23$^\star$ &   75.7 & 	 &    7001.55$^\star$ &   49.1 \\
	 &    6233.19$^\star$ &   62.5 & 	 &    6270.23~~ &   92.5 & 	 &    7062.97~~ &   57.9 \\
	 &    6266.30~~ &   47.9 & 	 &    6380.75~~ &   76.8 & 	 &    7715.59$^\star$ &   78.7 \\
\tablerule
\end{tabular}
\vskip2mm
\parbox[c]{120mm}{\baselineskip=9pt $\star$~The asterisk indicates the
averaged equivalent widths from the spectra of 1999 and 2006, while
other equivalent widths are only from the spectrum of 1999.} }
\end{table}

In the spectra we selected 135 atomic lines for the measurement of
equivalent widths and 17 lines for the comparison with synthetic
spectra.  The measured equivalent widths of lines are presented in
Table~1, in overlapping regions they show a very good agreement
(see Figure~1) and were averaged.

\subsectionb{2.1}{Atmospheric parameters}

Initially, the effective temperature, $T_{\rm eff}$, for 33~Psc was
estimated and averaged from the intrinsic color indices $(B-V)_0$ and
$(b-y)_0$, using the calibrations by Alonso et al.\ (1999).
The values of color indices, $B-V=1.06$ and $b-y=0.627$, were taken from
Reglero et al.\ (1987) and Hauck \& Mermilliod (1998), respectively.  A
small dereddening correction of $E_{B-V}=0.01$, estimated using the
Hakkila et al.\ (1997) software, was taken into account.

The agreement between the temperatures deduced from the two color
indices was quite good, with a difference of 80~K only.  No obvious
trend of the Fe\,{\sc i} abundances with the excitation potential was
found (Figure~2).

The surface gravity $\log\,g$ was found by adjusting the model gravity
to yield the same iron abundance from the Fe\,{\sc i} and Fe\,~{\sc ii}
lines.  The microturbulent velocity $v_{\rm t}$  corresponding to a
minimal line-to-line Fe\,{\sc i} abundance scattering was chosen as the
correct value.  Consequently, [Fe/H] values do not depend on the
equivalent widths of lines (Figure~3).

\vspace{3mm}
%--------------------- Fig. 2
\vbox{
\centerline{\psfig{figure=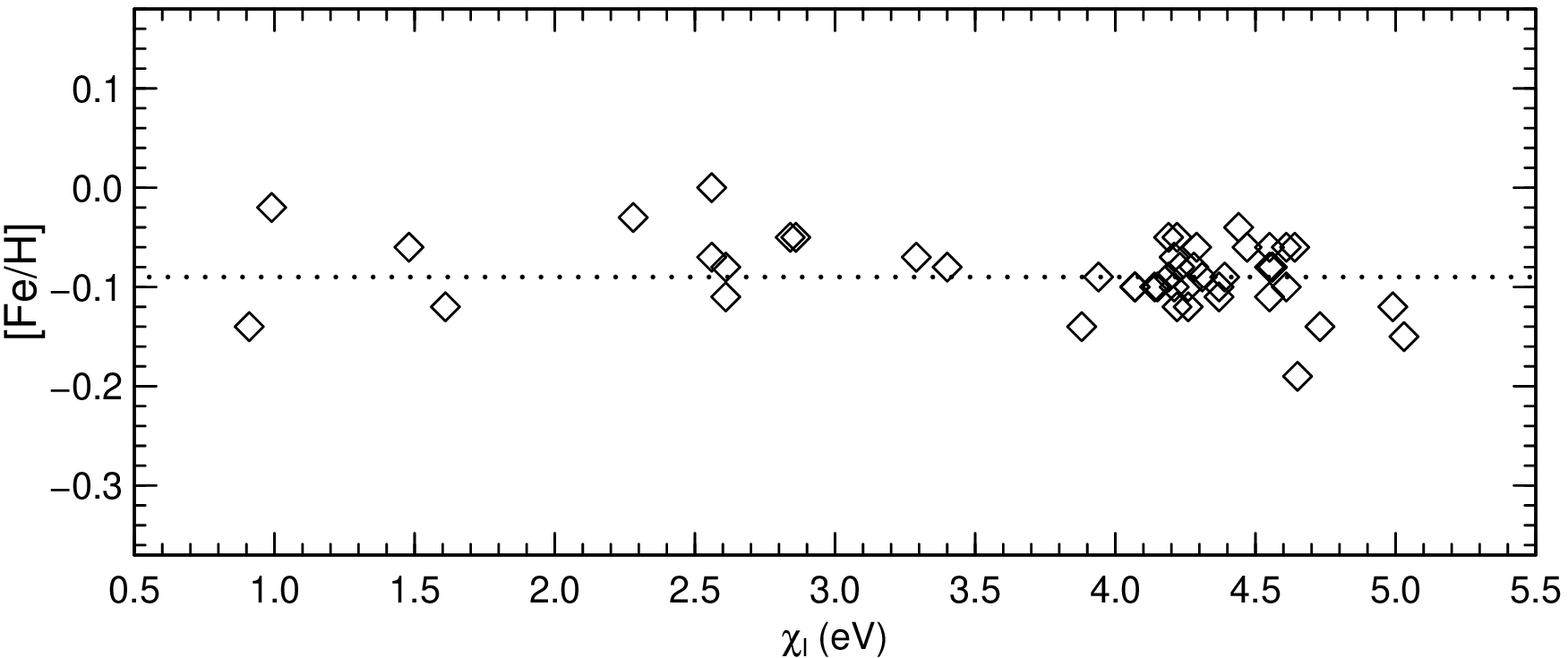,width=115truemm,angle=0,clip=}}
\captionb{2}{The [Fe\,{\sc i}/H] abundance values versus the lower
excitation potential $\chi_{\rm l}$.  The mean abundance
([Fe\,{\sc i}/H] = --0.09~dex) is shown as a dotted line.}
}
\vspace{3mm}

%-------------------- Fig. 3
\vbox{
\centerline{\psfig{figure=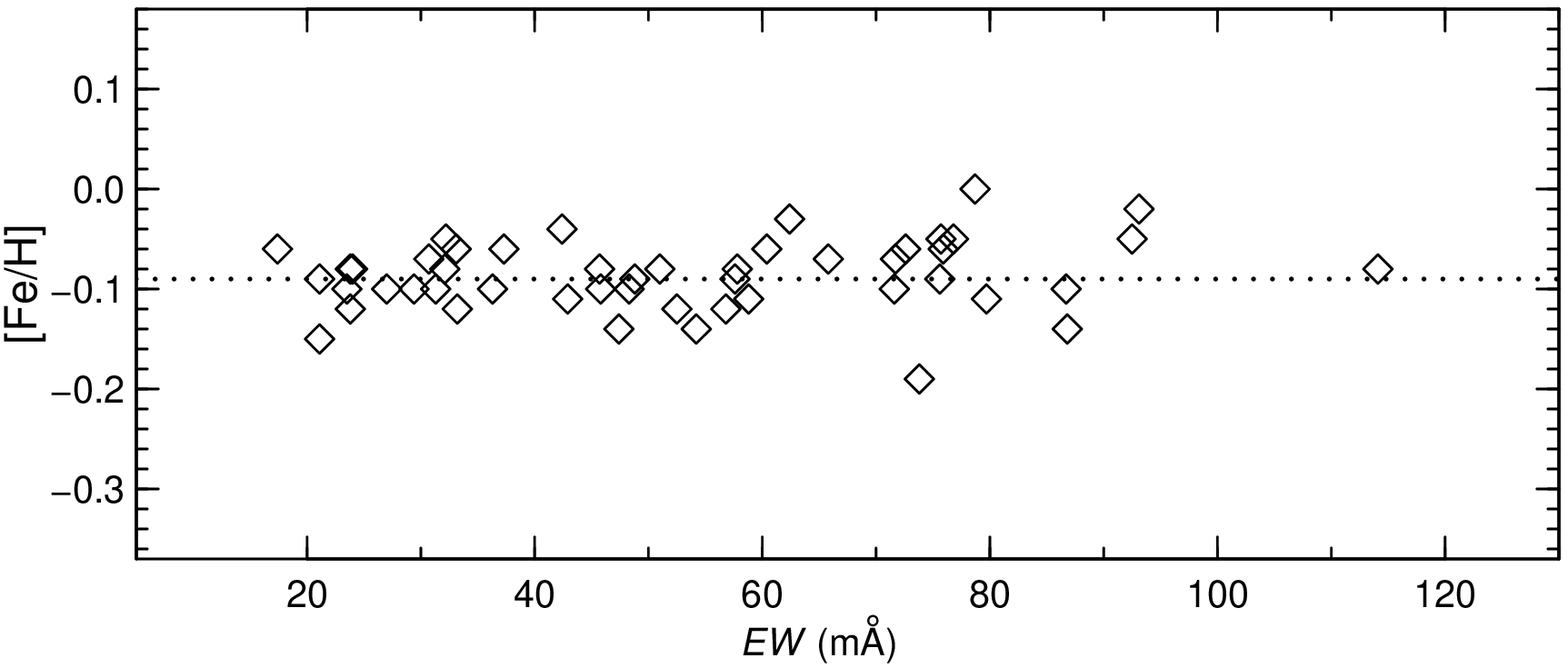,width=115truemm,angle=0,clip=}}
\captionb{3}{The [Fe\,{\sc i}/H] abundance values versus
the equivalent widths. The mean abundance ([Fe\,{\sc
i}/H] = --0.09~dex) is shown as a dotted line.}
}
\vspace{3mm}

\subsectionb{2.2}{Mass determination}

The mass of 33~Psc was evaluated from its effective temperature,
luminosity and the isochrones from Girardi et al.\ (2000).  The
luminosity $\log\,(L/L_{\odot})=1.39$ was calculated from the {\it
Hipparcos} parallax $\pi=25.32$~mas and $V = 4.61$~mag (van Leeuwen
2007), the bolometric correction calculated according to Alonso et al.\
(1999), and the above mentioned $E_{B-V}=0.01$.  The mass of 33~Psc
$\sim$\,1.6\,$M_{\odot}$ was found.

Previous mass determinations for 33~Psc are within 1.1\,$M_{\odot}$
(Gottlieb \& Bell 1972) and 3.0\,$M_{\odot}$ (Barrado y Navascues et
al.\ 1998).  The masses 1.6\,$M_{\odot}$ obtained by Glebocki \&
Stawikowski (1979) and 1.7\,$M_{\odot}$ by Pourbaix \& Boffin (2003) and
Morel et al.\ (2004) are close to our result.

\vspace{3mm}
\subsectionb{2.3}{Spectrum syntheses}

The method of synthetic spectra was used to determine carbon abundance
from the C$_2$ line at 5135.5~{\AA}.  The interval 7980--8130~{\AA},
containing strong $^{12}$C$^{14}$N and $^{13}$C$^{14}$N features, was
used for the nitrogen abundance and the $^{12}$C/$^{13}$C ratio
determinations.  The $^{12}$C/$^{13}$C ratio was determined from the
(2,0) $^{13}$C$^{12}$N feature at 8004.7~{\AA}.  All log\,$gf$ values
were calibrated to fit to the solar spectrum by Kurucz (2005) with solar
abundances from Grevesse \& Sauval (2000).

The oxygen abundance was determined from the forbidden [O\,{\sc i}] line
at 6300.31~\AA\ with the oscillator strength values for $^{58}$Ni and
$^{60}$Ni from Johansson et al.  (2003) and the values log~$gf = -9.917$
obtained by fitting to the solar spectrum (Kurucz 2005) and
log~$A_{\odot}=8.83$ taken from Grevesse \& Sauval (2000).

In Figures~4 and 5 we show several examples of synthetic spectra in the
vicinity of C$_2$, [O{\sc i}] and $^{12}$C$^{14}$N lines.

\vspace{3mm}
%--------------------------------------------------------------------------- Fig. 4
\vbox{
\centerline{\psfig{figure=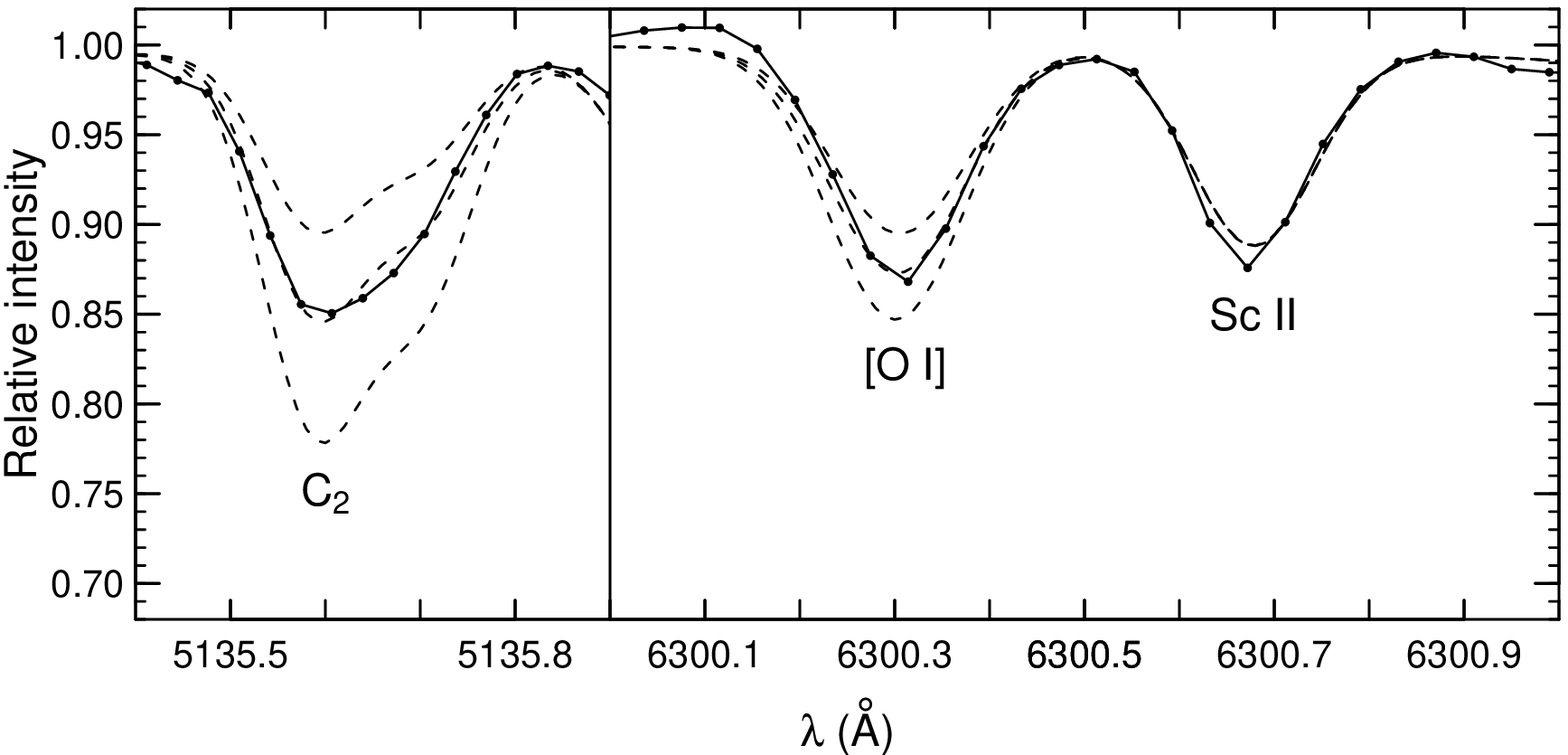,width=125truemm,angle=0,clip=}}
\vspace{-1mm}
   \captionb{4}{Synthetic spectrum fits to the C$_2$ line at
5135.5~{\AA} and
the forbidden [O\,{\sc i}] line at 6300.3~\AA. The observed spectra
are shown as solid lines. The  dashed lines are synthetic spectra with
${\rm [C/Fe]} =-0.03$, 0.13 and 0.23~dex and ${\rm [O/Fe]} = 0.03$,
0.13 and 0.23~dex.}
}
\vspace{3mm}

%------------------------------------------------------------------------------- Fig. 5
\vbox{
\centerline{\psfig{figure=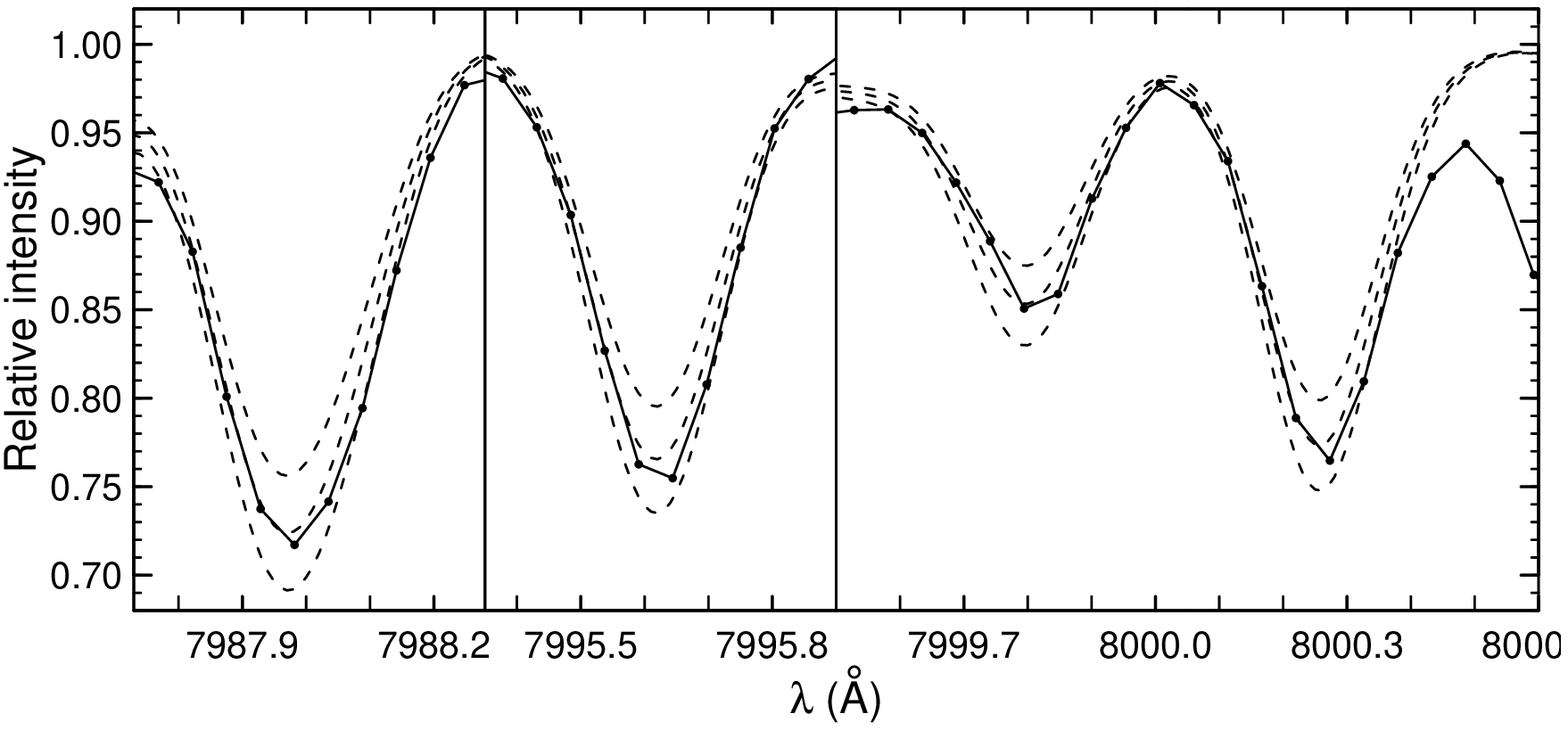,width=125truemm, angle=0,clip=}}
\vspace{-1mm}
\captionb{5}{Synthetic spectrum fits to $^{12}$CN lines.
The observed spectrum is shown as a solid line. The dashed lines are
synthetic spectra with ${\rm [N/Fe]} = 0.13$, 0.23 and 0.33~dex.}
}
\vspace{3mm}

%-------------------------------------------------------------------------------- Fig. 6
\vbox{
\centerline{\psfig{figure=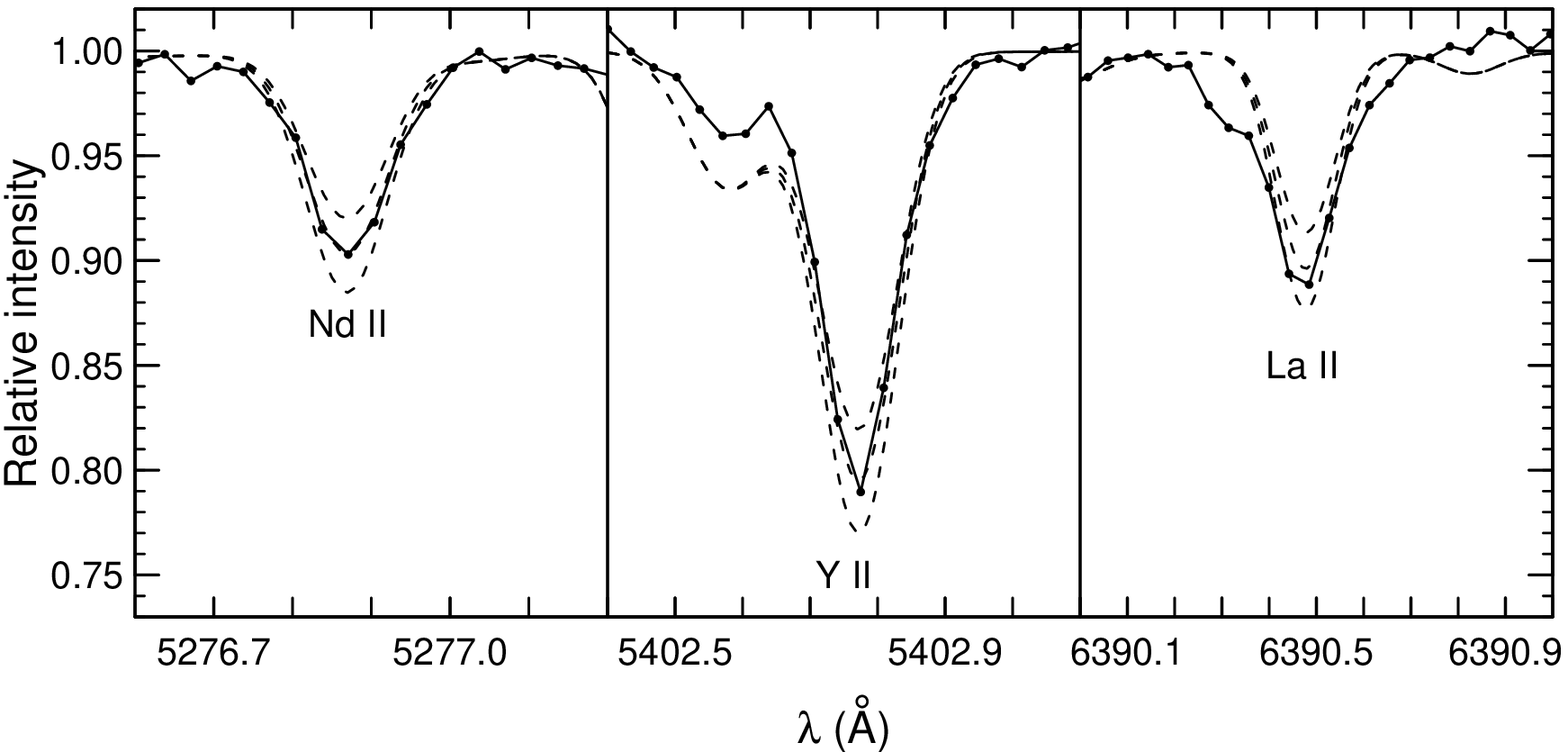,width=125truemm,angle=0,clip=}}
\vspace{-1mm}
\captionb{6}{Synthetic  spectrum fit to the Nd\,{\sc ii} line at
5276.806~\AA,  Y\,{\sc ii} at 5402.78~\AA\  and  La\,{\sc ii} at
6390.48~\AA.  The observed spectrum is shown as a solid
line. The  dashed lines are synthetic spectra  with
${\rm [Nd/Fe]} = 0.11$, 0.21 and 0.31~dex, ${\rm [Y/Fe]} = -0.07$, 0.03
and
0.13~dex and ${\rm [La/Fe]} = -0.07$, 0.03 and 0.13~dex
for Nd\,{\sc ii}, Y\,{\sc ii} and La\,{\sc ii} lines, respectively.}}
\vspace{2mm}

The abundance of Na\,{\sc i} was estimated using the line at
5148.84~\AA\ which is slightly blended by the Ni\,{\sc i} 5148.66~\AA
line.  These two lines are distinct in the Sun, so we were able to
calibrate their log\,$gf$ values using the solar spectrum.  So, the
sodium abundance value in our study is affected by uncertainty of the
nickel abundance determination by the Equivalent Widths method.
Fortunately, the line-to-line scatter of [Ni/H] determination from 20
lines of Ni\,{\sc i} was as small as 0.06~dex.

For the evaluation of Zr\,{\sc i} abundance the lines at 5385.13~\AA,
6127.48~\AA\ and 6134.57~\AA\ were used.  Evaluation of the Y\,{\sc ii}
abundance was based on the 5402.78~\AA\ line, La\,{\sc ii} on the
6390.48~\AA\ line, Ce\,{\sc ii} on the 5274.22~\AA\ and 6043.38~\AA\
lines and Nd\,{\sc ii} on the 5276.86~\AA\ line.  The synthetic spectrum
fits to the Nd\,{\sc ii} line at 5276.806~\AA, Y\,{\sc ii} at
5402.78~\AA\ and La\,{\sc ii} at 6390.48~\AA\ are displayed in Figure~6.

The abundance of $r$-process element praseodymium was based on the
Pr\,{\sc ii} line at 5259.72~\AA\ and of europium on the Eu\,{\sc ii}
line at 6645.10~\AA\ (Figure~7).  The hyperfine structure of Eu\,{\sc
ii} was taken into account when calculating the synthetic spectrum.  The
wavelength, excitation energy and total log~$gf = 0.12$ were taken from
Lawler et al.\ (2001), the isotopic meteoritic fractions of $^{151}{\rm
Eu}$, 47.77\%, and $^{153}{\rm Eu}$, 52.23\%, and isotopic shifts were
taken from Biehl (1976).

Due to slow rotation spectral lines of 33~Psc are broadened very little.
We used $v\,{\rm sin}\,i=1.9~{\rm km\,s}^{-1}$ from Batten et al.\
(1989), which fits well the profiles of spectral lines of 33~Psc.

\vspace{2mm}
%------------------------------------------------------------------------------- Fig. 7
\vbox{
\centerline{\psfig{figure=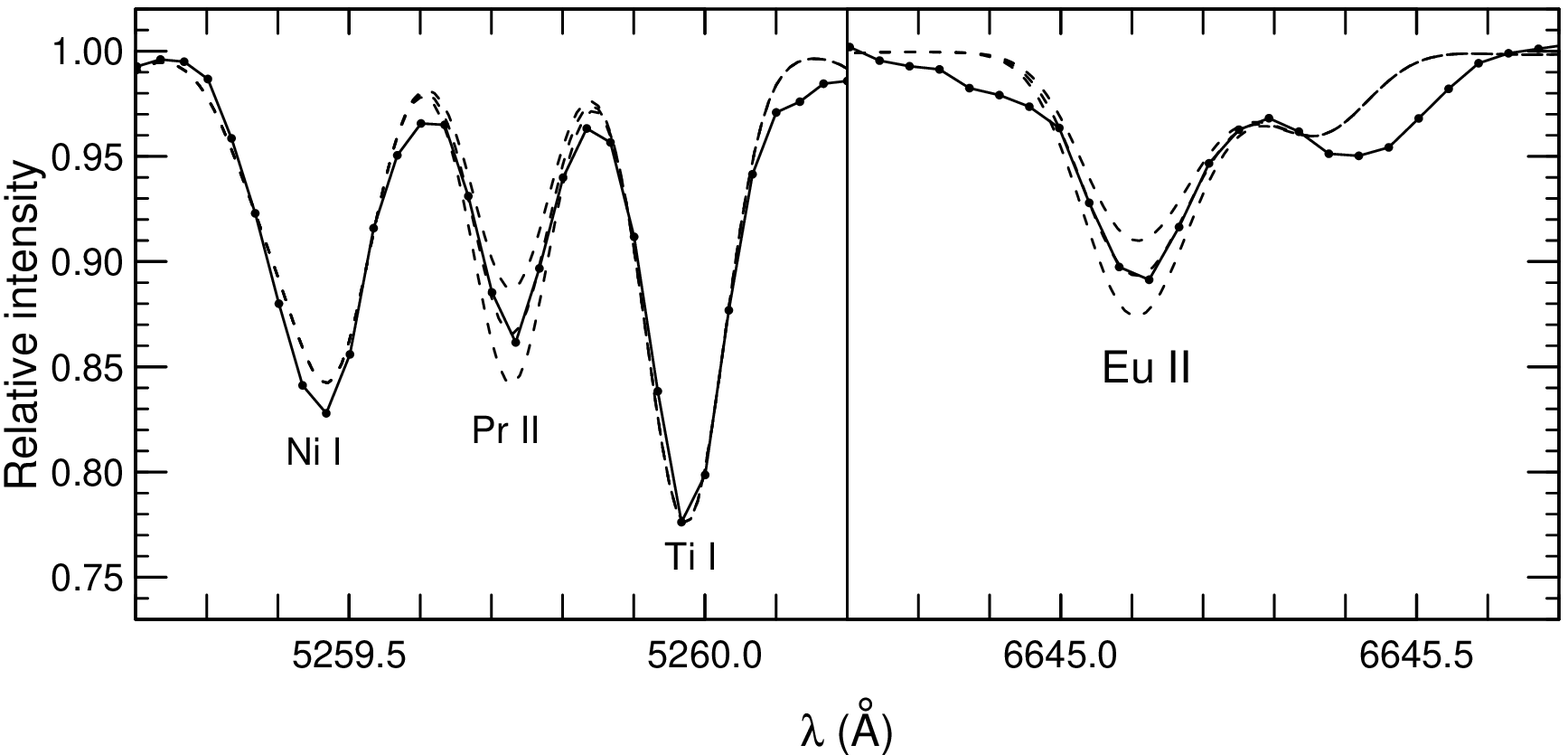,width=125truemm,angle=0,clip=}}
\vspace{-2mm}
\captionb{7}{Synthetic spectrum fit to the Eu\,{\sc ii}  line at
6645.10~\AA. The observed spectrum
is shown as a solid line. The dashed lines  are synthetic spectra with
${\rm [Eu/Fe]} = 0.02$, 0.12 and 0.22~dex.}
}
\vspace{2mm}

\subsectionb{2.4}{Estimation of uncertainties}

The sources of uncertainty were described in detail in Paper~I.
The sensitivity of the abundance estimates to changes in the atmospheric
parameters for the assumed errors ($\pm~100$~K for $T_{\rm eff}$, $\pm
0.3$~dex for log~$g$ and $\pm 0.3~{\rm km~s}^{-1}$ for $v_{\rm t}$) is
illustrated in Table~2.  It is seen that possible parameter errors do
not affect the abundances seriously; the element-to-iron ratios, which
we use in our discussion, are even less sensitive.

The scatter of the deduced line abundances $\sigma$, presented in
Table~3, gives an estimate of the uncertainty due to the random errors,
e.g., in the continuum placement and the line parameters (the mean value
of $\sigma$ is 0.06~dex).  Thus, the uncertainties in the derived
abundances originating from the random errors are close to this value.

Since the abundances of C, N and O are bound together by the molecular
equilibrium, we have also investigated how the error in one of them
typically affects the abundance determination of others.

$\Delta{\rm [O/H]}=0.10$ causes
$\Delta{\rm [C/H]}=0.04$ and $\Delta{\rm [N/H]}=0.05$;
$\Delta{\rm [C/H]}=0.10$ causes $\Delta{\rm [N/H]}=-0.12$ and
$\Delta{\rm [O/H]}=0.02$. $\Delta {\rm [N/H]}=0.10$ has no effect
on both carbon and oxygen abundances.

\vspace{3mm}
%----------------------- Table 2
\begin{center}
\vbox{\small
\tabcolsep=3pt
\begin{footnotesize}
\parbox{120mm}{\baselineskip=9pt
{\small \bf \ \ Table 2.~}{\small Sensitivity
of the abundances to changes of the atmospheric parameters.
The table  entries show the effects on the logarithmic abundance
relative to hydrogen, $\Delta$\,[El/H].
}}
\begin{tabular}{lD..{1.1}D..{1.1}D..{1.1}cD..{1.1}D..{1.1}D..{1.1}}
\noalign{\vskip0.5mm}
\tablerule
\multicolumn{1}{l}{Ion}&
\multicolumn{1}{c}{$\Delta T_{\rm eff}$}&
\multicolumn{1}{c}{$\Delta {\rm log}~g$}&
\multicolumn{1}{c}{$\Delta v_{\rm t}$}&

\multicolumn{1}{c}{Ion }&
\multicolumn{1}{c}{$\Delta T_{\rm eff}$}&
\multicolumn{1}{c}{$\Delta {\rm log}~g$}&
\multicolumn{1}{c}{$\Delta v_{\rm t}$}\\

\multicolumn{1}{c}{}&
\multicolumn{1}{c}{$+100$~K}&
\multicolumn{1}{c}{$+0.3$}&
\multicolumn{1}{c}{$+0.3 {\rm km~s}^{-1}$}&

\multicolumn{1}{c}{}&
\multicolumn{1}{c}{$+100$~K}&
\multicolumn{1}{c}{$+0.3$}&
\multicolumn{1}{c}{$+0.3 {\rm km~s}^{-1}$}\\
\tablerule
\noalign{\vskip0.3mm}
C(C$_2$)& 	0.02  &  	0.11	  &  	0.00	  &	Fe\,{\sc ii}	& 	-0.09	  &  	0.17	  &  	-0.06	 \\
N(CN)& 	0.05	  &  	0.13	  &  	0.00	  &	Co\,{\sc i}		& 	0.05	  &  	0.06	  &  	 -0.08	 \\
O([O\,{\sc i}])	& 	0.00  &  0.13  &  	0.00	  &	Ni\,{\sc i}		& 	0.01	  &  	0.06	  &  	 -0.10	 \\
Na\,{\sc i}	& 0.08  &    -0.01	  &  	0.00	  &	Y\,{\sc ii}		& 	0.00	  &  	0.11	  &  	 -0.03	 \\
Si\,{\sc i}	& -0.04	  &  	0.06  &  	-0.04	  &	Zr\,{\sc i}		& 	0.20	  &    -0.04	  &  	 -0.01	 \\
Ca\,{\sc i}	& 	-0.09  &    -0.01	  & -0.08	  &	La\,{\sc ii}	& 	0.01	  &  	0.13	  &  	 -0.01	 \\
Sc\,{\sc ii}& 	-0.02	  &  	0.12  & -0.09	  &	Ce\,{\sc ii}& 	0.00	  &  	0.08	  &  	 0.00	 \\
Ti\,{\sc i}	& 0.12  &  	0.01	  &  	-0.08	  &	Pr\,{\sc ii}	& 	0.01	  &  	0.13	  &  	 -0.01	 \\
V\,{\sc i}	& 0.15  &  	0.01	  &  	-0.10	  &	Nd\,{\sc ii}	& 	0.01	  &  	0.13	  &  	 -0.01	 \\
Cr\,{\sc i}	& 0.08	  &    -0.01  &  	-0.08	  &	Eu\,{\sc ii}	& 	-0.01	  &  	0.13	  &  	 0.00	 \\
Fe\,{\sc i}	& 	0.04  &  0.04	  &  	-0.08	  &		&		&		&		\\
\noalign{\vskip1mm}
C/N & 	-0.23  &  -0.10  &  0.10  &
\multicolumn{1}{l}{\textsuperscript{12}C/\textsuperscript{13}C} &  	
\multicolumn{1}{c}{--2} &  \multicolumn{1}{c}{1}	&\multicolumn{1}{c}{--2}	   \\
\tablerule
\end{tabular}
\end{footnotesize}
}
\end{center}

\vspace{1mm}

%---------------------------------- Table 3

\begin{center}
\vbox{\small
\tabcolsep=3pt
\begin{footnotesize}
\parbox{120mm}{%\baselineskip=12pt
 {\small \bf \ \ Table 3.}{\small \ Element abundances relative to
hydrogen, [El/H]. $\sigma$ is a standard deviation in the
mean value due to the line-to-line scatter within the species. $N$ is
the number of lines used for the abundance determination.
 }}
\begin{tabular}{lrrclrrc}
\noalign{\vskip0.5mm}
\tablerule
\multicolumn{1}{l}{Ion }&
\multicolumn{1}{c}{$N$}&
\multicolumn{1}{l}{[El/H]}&
\multicolumn{1}{c}{$\sigma$ }&
\multicolumn{1}{l}{Ion }&
\multicolumn{1}{c}{$N$}&
\multicolumn{1}{c}{[El/H]}&
\multicolumn{1}{c}{$\rm \sigma$ } \\
\noalign{\vskip0.3mm}
\tablerule
\noalign{\vskip0.3mm}
C(C$_2$)				& 1		& $-0.13$	& $-$  		& Fe\,{\sc ii}			& 6		& $-0.08$	& $0.06$ \\
N(CN)$^\star$			& 4		& $0.14$	& $0.02$  	& Co\,{\sc i}			& 5		& $0.05$	& $0.07$ \\
O([O\,{\sc i}]) 		& 1		& $0.04$	& $-$  		& Ni\,{\sc i}$^\star$	& 20	& $ 0.00$	& $0.06$ \\
Na\,{\sc i}				& 1		& $-0.01$	& $-$  		& Y\,{\sc ii}			& 1		& $-0.06$	& $-$ \\
Si\,{\sc i}$^\star$		& 6		& $0.04$	& $0.06$  	& Zr\,{\sc i}			& 3		& $-0.21$	& $0.08$ \\
Ca\,{\sc i}$^\star$		& 4		& $-0.02$	& $0.07$  	& La\,{\sc ii}$^\star$	& 1		& $-0.06$ 	& $-$ \\
Sc\,{\sc ii}			& 4		& $0.13$	& $0.08$  	& Ce\,{\sc ii}			& 2		& $-0.12$	& $0.07$ \\
Ti\,{\sc i}$^\star$		& 18	& $-0.01$	& $0.06$  	& Pr\,{\sc ii}			& 1		& $-0.02$	& $-$ \\
V\,{\sc i}$^\star$		& 16	& $0.07$	& $0.07$  	& Nd\,{\sc ii}			& 1		& $0.12$	& $-$ \\
Cr\,{\sc i}				& 7		& $0.01$	& $0.05$  	& Eu\,{\sc ii}			& 1		& $0.03$	& $-$ \\
Fe\,{\sc i}$^\star$		& 49	& $-0.09$	& $0.04$  		& 	&	&	&	\\
%\\
%C/N              & 5 & $2.14$ & 0.27 & \textsuperscript{12}C/\textsuperscript{13}C & 1 & 15 & $-$\\
\tablerule
\end{tabular}
\\
\vspace{2mm}
\parbox{120mm}{$\star$~The asterisk indicates the elements, whose
abundances
were calculated from the 1999 and 2006 spectra, while
the others were calculated only from the 1999 spectrum.}
\end{footnotesize}
}
\end{center}
%---------------------------------------------------------------------------------------------------

\sectionb{3}{RESULTS AND DISCUSSION}

The following atmospheric parameters are obtained for 33~Psc:
$T_{\rm eff}=4750$~K,
${\rm log}\,g=2.8$, $v_{\rm t}=1.1~{\rm km}\,s^{-1},$
${\rm [Fe/H]}=-0.09$,
${\rm [C/Fe]} = -0.04$, ${\rm [N/Fe]} = 0.23$, ${\rm [O/Fe]} = 0.13$,
as well as the ratios C/N = 2.14 and $^{12}$C/$^{13}$C = 30.   The
element abundances [A/H] and the
$\sigma$ values (the line-to-line scatter) are listed in Table~3.

In Figure 8 the obtained values of [El/Fe] for 33~Psc are compared with
the results of other authors.  Although McWilliam (1990) has used a
high-resolution spectrum, his values of [El/Fe] are rather
scattered, probably due to small number of lines measured.  The recent
results of Morel et al.\ (2004) are in good agreement with ours.  The
element to iron ratios in 33~Psc are close to solar.

\vspace{3mm}
%-------------------------------------------------------------------------------- Fig. 8
\vbox{
\centerline{\psfig{figure=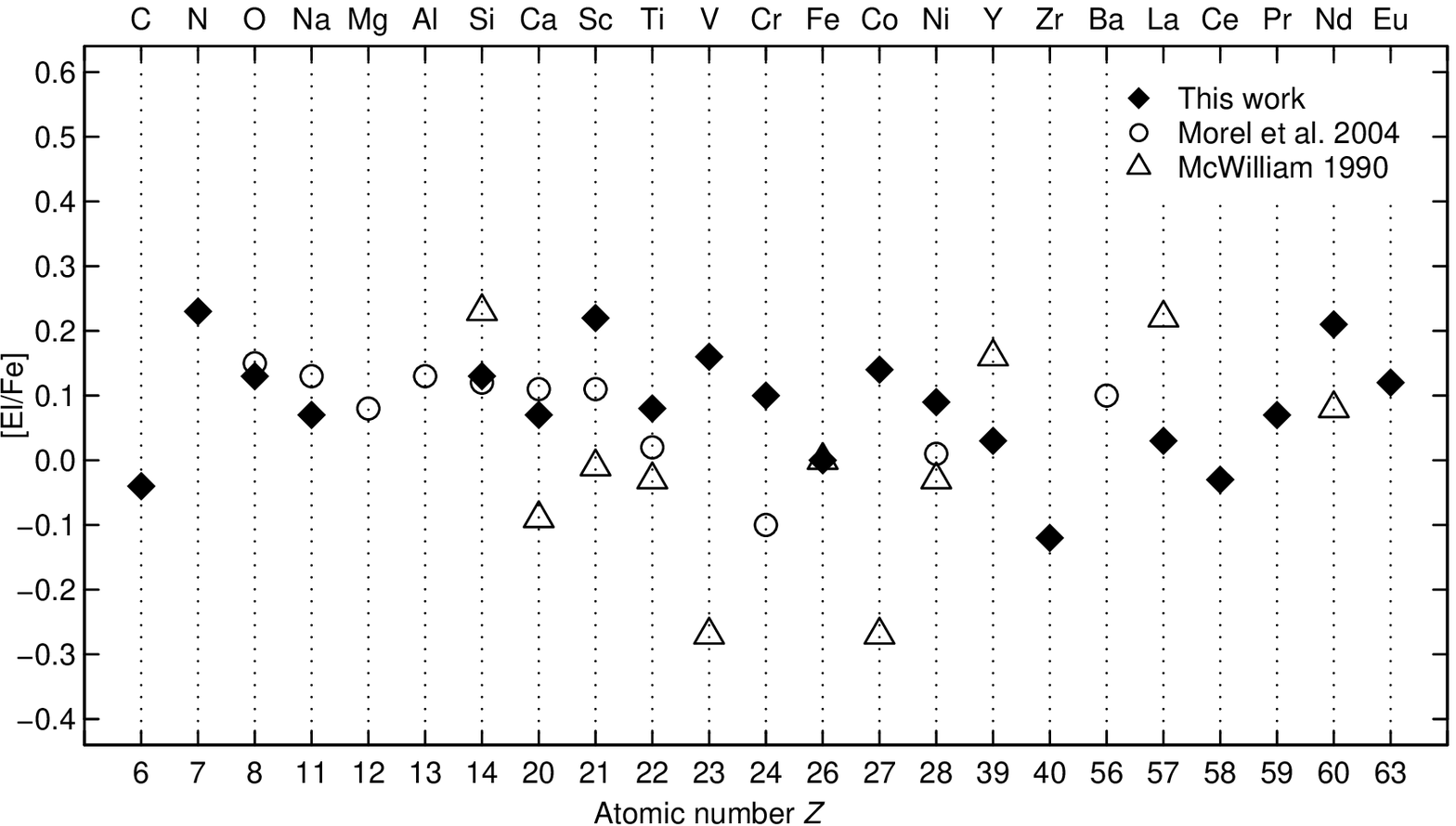,width=120mm, angle=0,clip=}}
\vspace{-1mm}
\captionb{8}{Element abundances for 33~Psc, as determined in this work
(filled diamonds), by Morel et al.\ (2004, circles)  and by McWilliam
(1990, triangles).}
}
\vspace{4mm}
%--------------------------------------------------------------------------------- Fig. 9
\vbox{
\centerline{\psfig{figure=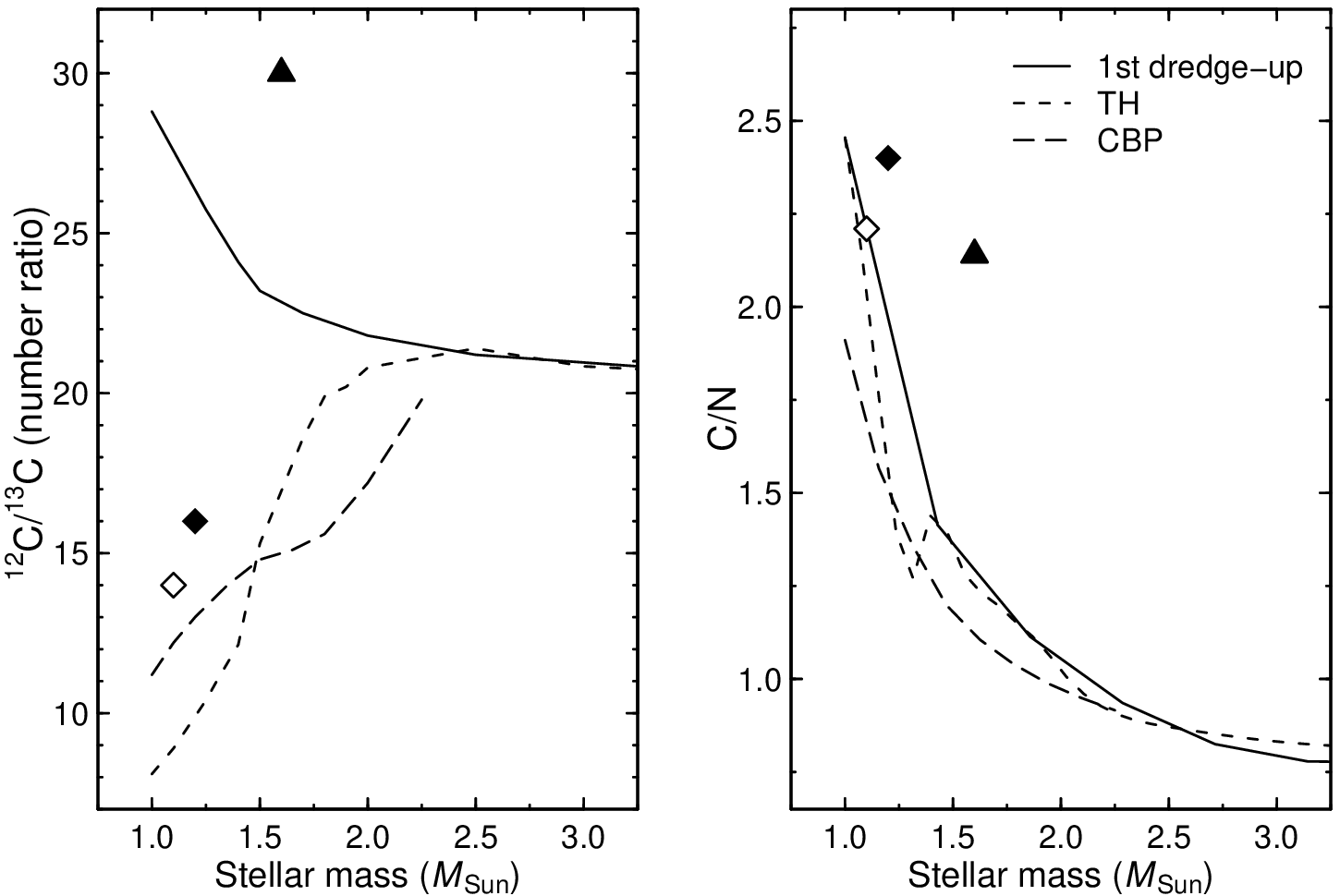,width=120truemm,angle=0,clip=}}
\vspace{-1mm}
  \captionb{9}{Comparisons of C/N and $^{12}$C/$^{13}$C ratios in
33~Psc (filled triangles),
$\lambda$~And (open diamonds, Paper~I) and 29~Dra (filled diamonds,
Paper II) with the theoretical predictions explained in the text.}
}
\vspace{3mm}

The evolutionary sequences in the luminosity vs. effective
temperature diagram by Girardi et al.\ (2000) show that 33~Psc with its
luminosity ${\rm log}(L/L_{\odot})=1.39$ is a first ascent giant lying
slightly below the red giant sequence bump at ${\rm
log}\,(L/L_{\odot})=1.6$ (Charbonnel \& Lagarde 2010).  According to the
mentioned model of mixing, carbon and nitrogen abundances in 33~Psc
should be altered only by the first dredge-up.  Our results confirm this
prediction.  In Figure~9 we compare C/N and $^{12}$C/$^{13}$C ratios of
33~Psc with the theoretical models including extra-mixing.  The model,
called `cool bottom processing' (CBP), was proposed by Boothroyd \&
Sackmann (1999) and the model, called `thermohaline mixing' (TH), was
proposed by Charbonnel \& Lagarde (2010).  The position of 33~Psc in
Figure~9 indicates that its ratios of carbon isotopes and C/N have not
been altered by extra mixing.

Two more active RS CVn stars $\lambda$~And and 29~Dra, investigated by
now in our program, gave a hint that extra-mixing processes may start
acting in these low-mass chromospherically active fast rotating stars
slightly earlier than at the bump of the red giant sequence in
non-active stars.  The star 33~Psc with negligible activity seems to be
normal in this respect.

The abundance of lithium is also very sensitive to mixing.  During the
first dredge-up, for a star of the mass of 33~Psc, the Li abundance
drops to approximately log\,$A{\rm (Li)}=1.44$ (Charbonnel \& Lagarde
2010).  However, the available determinations of Li abundance in 33~Psc
show lower abundances:  Brown et al.\ (1989) found log\,$A{\rm
(Li)}=0.8$, Pallavicini et al.\ (1992) and Randich et al.\ (1994) found
log\,$A{\rm (Li)}\le 0.1$, and Barrado y Navascues et al.\ (1998) found
log\,$A{\rm (Li)}=0.41$.  These values are lower than predicted by the
first dredge-up model, while the carbon and nitrogen abundances are in
agreement with it.  Thus, the lowered lithium abundance should be
related to other mechanisms.

\thanks{This project has been supported by the European Commission
through the Baltic Grid II project.}

\References

\refb Alonso~A., Arribas~S., Mart\'{i}nez-Roger~C. 1999, A\&AS, 140, 261

\refb Barrado~y~Navascues~D., de~Castro~E., Fernandez-Figueroa~M.~J., Cornide~M., Garcia~Lopez~R.~J. 1998,  A\&A, 337, 739

\refb Barisevi\v{c}ius~G., Tautvai\v{s}ien\.{e}~G., Berdyugina~S.,
Chorniy~Y., Ilyin~I. 2010, Baltic Astronomy, 19, 157 (Paper II)

\refb Basri G., Laurent R., Walter F. M. 1985, ApJ, 298, 761

\refb Batten~A.~H., Fletcher~J.~M., MacCarthy~D.~G., 1989, PDAO, 17, 1

\refb Biehl~D. 1976, Diplomarbeit, Christian-Albrechts-Universit\"at
Kiel, Institut f\"ur Theoretische Physik und Sternwarte

\refb Boothroyd~A.~I., Sackman~I.~J. 1999, ApJ, 510, 232

\refb Brown J. A., Sneden C., Lambert D. L., Dutchover E. Jr. 1989, ApJS, 71, 293

\refb Charbonnel C., Lagarde N. 2010, A\&A, 522, A10

\refb Drake S. A., Simon T. Linsky J. L. 1989, ApJS, 71, 905

\refb Girardi L., Bressan A., Bertelli G., Chiosi C. 2000, A\&AS, 141, 371

\refb Glebocki~R., Stawikowski~A. 1979, AcA, 29, 505

\refb Gottlieb~D.~M., Bell~R.~A. 1972, A\&A, 19, 434

\refb Grevesse N., Sauval A. J. 2000, in {\it Origin of Elements in the
Solar System, Implications of Post-1957 Observations}, ed. O. Manuel,
Kluwer, p.\,261

\refb Hakkila J., Myers J. M., Stidham B. J., Hartmann D. H. 1997, AJ, 114, 2043

\refb Hansen~L., Kjaergaard~P. 1971, A\&A, 15, 123

\refb Harper W. E. 1926, Publ. Dominion Astrophys. Obs., 3, 341

\refb Hartkopf W. I., McAlister H. A., Mason B. D. 2001, AJ, 122, 3480

\refb Hauck~B., Mermilliod~M. 1998, A\&AS, 129, 431

\refb Johansson~S., Litzen~U., Lundberg~H., Zhang~Z. 2003, ApJ, 584, 107

\refb Kurucz~R. L. 2005, {\it New Atlases for Solar Flux, Irradiance,
Central Intensity, and Limb Intensity}, Mem. SA Ital. Suppl., 8, 189

\refb Lawler J. E., Wickliffe M. E., Den Hartog E. A. 2001, ApJ, 563, 1075

\refb McWilliam A. 1990, ApJS, 74, 1075

\refb Morel~T., Micela~G., Favata~F., Katz~D. 2004, A\&A, 426, 1007

\refb Pallavicini R., Randich S., Giampapa M. S. 1992, A\&A, 253, 185

\refb Pourbaix~D., Boffin~H.\,M.\,J. 2003, A\&A, 398, 1163

\refb Randich~S., Giampapa~M.~S., Pallavicini~R. 1994, A\&A, 283, 893

\refb Reglero~V., Gimenez~A., de Castro~E., Fernandez-Figueroa~M. J., 1987, A\&AS, 71, 421

\refb Strassmeier~K.~G., Hall~D.~S., Zeilik~M., Nelson~E., Eker~Z., Fekel~F.~C. 1988, A\&AS, 72, 291

\refb Tautvai\v{s}ien\.{e}~G., Barisevi\v{c}ius~G., Berdyugina~S.,
Chorniy~Y., Ilyin~I. 2010, Baltic Astronomy, 19, 95 (Paper~I)

\refb van Leeuwen~F. 2007, {\it Hipparcos, the New Reduction of the Raw
Data}, Astrophysics and Space Science Library, vol. 350

\refb Walter~F.~M. 1985, PASP, 97, 643

\refb Young A., Konigest A. 1977, ApJ, 211, 836

\end{document}